\documentclass[12pt]{article}
\usepackage{amssymb,epsfig,psfig}

\newcommand{\bn}{{\bf n}}
\newcommand{\bu}{{\bf u}}
\newcommand{\bw}{{\bf w}}

\begin{document}

\title{Dodecahedral topology fails to explain quadrupole-octupole alignment}

\author{Jeff Weeks\\Jesper Gundermann}

\maketitle

\begin{abstract}
    \noindent The CMB quadrupole and octupole,
    as well as being weaker than expected,
    align suspiciously well with each other.
    Non-trivial spatial topology can explain
    the weakness.  Might it also explain the alignment?
    The answer, at least in the case of the
    Poincar\'e dodecahedral space, is a resounding no.
\end{abstract}

\section{Introduction}
\label{SectionIntroduction}

Soon after the release of the first-year WMAP
data~\cite{WMAP-Bennett}, Tegmark {\it et al.} ~\cite{Tegmark}
noticed that the CMB quadrupole and octupole aligned with each
other unusually well, at roughly the $98\%$ level. Multipole
vectors -- discovered by Maxwell~\cite{Maxwell} in the
$19^\mathrm{th}$ century, widely forgotten, then reintroduced by
Copi {\it et al.}~\cite{Copi} -- provide a useful tool for
analyzing the alignment in greater detail.  While exact confidence
levels vary depending on what one measures, all researchers agree
that the quadrupole-octupole alignment is unusual at roughly the
$99\%$ level or better~\cite{Schwarz,Weeks,Bielewicz}.  The
combination of the 1-in-100 alignment with the 1-in-600 overall
weakness of the low-$\ell$ modes motivates one to seek a physical
explanation.

Non-trivial spatial topology can explain the weakness of the
low-$\ell$ modes.  Might it also explain the quadrupole-octupole
alignment?  The present paper simulates the CMB in a Poincar\'e
dodecahedral space~\cite{WeberSeifert33,Nature} and checks the
quadrupole-octupole alignment. Absolutely no correlation is found.

\section{Simulating the space}
\label{SectionSimulating}

We use the late Jesper Gundermann's simulation~\cite{Gundermann}
of the CMB in the Poincar\'e dodecahedral space, with modes
through $k_\mathrm{max} = 102$. This simplified simulation, while
neglecting the Doppler contribution and the sound speed,
nevertheless produces a low-$\ell$ power spectrum essentially
identical to the spectra produced by more refined simulations.
Thus we may be quite confident that if the dodecahedral topology
imposed a nontrivial quadrupole-octupole alignment, this
simulation would capture it. As we will see in
Section~\ref{SectionMeasuring}, however, absolutely no such
correlation is found.  Even if one were to add a Doppler term and
sound speed to the simulation, the distribution in
Figure~\ref{FigureQuadOctHistogram} would change by at most a tiny
amount, not nearly enough to introduce a nontrivial
quadrupole-octupole correlation.

\section{Measuring the alignment}
\label{SectionMeasuring}

For each simulated CMB sky, we use the polynomial
method~\cite{Weeks} to compute the two quadrupole vectors
$\{\bu_{2,1}, \bu_{2,2}\}$ and the three octupole vectors
$\{\bu_{3,1}, \bu_{3,2}, \bu_{3,3}\}$. Following~\cite{Schwarz},
we take the cross product $\bw_2 = \bu_{2,1} \times \bu_{2,2}$,
which we normalize to obtain a unit vector $\bn_2 = \bw_2/|\bw_2|$
orthogonal to the plane of the quadrupole. Similarly, we take the
cross product of each of the three possible pairs of octupole
vectors
\begin{eqnarray}
  \bw_{3,1} = \bu_{3,2} \times \bu_{3,3}\nonumber\\
  \bw_{3,2} = \bu_{3,3} \times \bu_{3,1}\nonumber\\
  \bw_{3,3} = \bu_{3,1} \times \bu_{3,2}
\end{eqnarray}
which we normalize to obtain unit vectors $\bn_{3,i} =
\frac{\bw_{3,i}}{|\bw_{3,i}|}$ orthogonal to each of the three
octupole planes.  The three dot products $D_i = | \bn_2 \cdot
\bn_{3,i} |$ then measure the extent to which the quadrupole plane
does or does not align with each of the three octupole planes.

In a simply connected universe one expects no correlation between
the quadrupole vector $\bn_2$ and each octupole vector
$\bn_{3,i}$.  As $\bn_2$ and $\bn_{3,i}$ (for some fixed $i$)
wander randomly over the 2-sphere, their dot product follows a
flat distribution on the interval $[-1,+1]$ (this is a consequence
of the wonderful fact that radial projection of a sphere onto a
circumscribed cylinder via $(x,y,z) \mapsto (\frac{x}{\sqrt{x^2 +
y^2}}, \frac{y}{\sqrt{x^2 + y^2}}, z)$ preserves area).  Hence
each $D_i = | \bn_2 \cdot \bn_{3,i} |$, being the absolute value
of the dot product, follows a flat distribution on $[0,1]$.

In the real universe the quadrupole aligns surprisingly well with
the octupole, giving dot products $\{D_1, D_2, D_3\} = \{0.84,
0.87, 0.95\}$ for the DQ-corrected Tegmark (DQT) cleaning
\cite{Tegmark} of the first-year WMAP data or $\{0.85, 0.87,
0.93\}$ for the Lagrange Internal Linear Combination (LILC)
cleaning \cite{Eriksen2} of the same data.

\begin{figure}
\centerline{\psfig{file=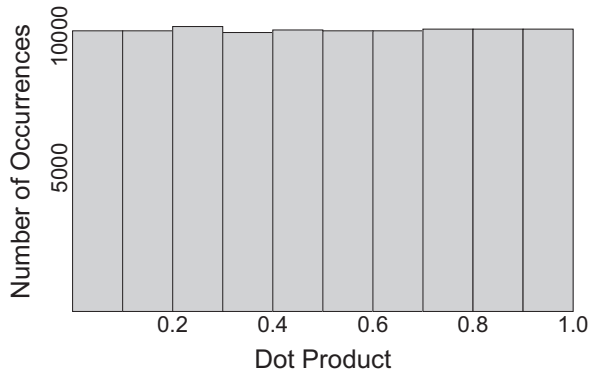, width=8cm}}
\caption{100000 simulations of the Poincar\'e dodecahedral space
find the distribution of the quadrupole-octupole dot product $|
\bn_2 \cdot \bn_{3,i} |$ to be completely flat, just as in a
simply connected universe. } \label{FigureQuadOctHistogram}
\end{figure}

The question of whether a multiconnected spatial topology might
explain the observed quadrupole-octupole alignment may be
rephrased more precisely as:  Does a given topology predict a flat
distribution for each $D_i$ or does it predict a distribution
skewed towards the high end?  For the Poincar\'e dodecahedral
space, our simulations (recall Section~\ref{SectionSimulating})
yield a flat distribution (Figure~\ref{FigureQuadOctHistogram}),
implying that the dodecahedral topology does nothing to explain
the quadrupole-octupole alignment.

To be fully rigorous we should point out that even though the
individual dot products $D_i$ follow the same flat distribution in
the dodecahedral topology that they do in the simply connected
model, it's nevertheless conceivable that their sum $D_1 + D_2 +
D_3$ might follow a slightly different distribution in the two
cases, depending on the internal correlations among the three
$D_i$ in the dodecahedral case.  In practice, however, our
simulations find the observed sum to be unusual at roughly the
$99\%$ level regardless of whether we compare to the dodecahedral
topology or a simply connected space.

\section{Conclusion}
\label{SectionConclusion}

The Poincar\'e dodecahedral space topology, while explaining the
weakness of the low-$\ell$ modes, completely fails to explain the
quadrupole-octupole alignment.  While this negative result leaves
one feeling less optimistic, good scientific practice demands that
one analyze a few other plausible topologies before reaching any
firm conclusion about whether topology might play a role.

One must also keep an open mind about what observations may or may
not be due to random chance alone.  The quadrupole-octupole
alignment might be due to chance, while the weakness of the
low-$\ell$ modes has a physical explanation.  Or perhaps exactly
the reverse is true.  At this point the mystery remains open.

\section*{Acknowledgments}
\label{SectionAcknowledgments}

I dedicate this article to my friend and collaborator Jesper
Gundermann, whose untimely death on 10 June 2006 saddened all who
knew him.  His energetic enthusiasm and deep love of science
brought joy to those of us lucky enough to work with him.

I thank the U.S. National Science Foundation for supporting this
work under grant DMS-0452612.

\end{document}